\documentclass[%
aps,
 reprint,
 amsmath,amssymb,
prper,
floatfix
]{revtex4-2}

\usepackage{graphicx}
\usepackage{dcolumn}
\usepackage{bm}
\usepackage{tikz}
\usepackage{pgfplots}
\usepackage[export]{adjustbox}
\pgfplotsset{compat=1.17}
\usepackage{hyperref}
\usepackage{xurl}

\begin{document}


\title{Tailoring Chatbots for Higher Education: Some Insights and Experiences}

\author{Gerd Kortemeyer}
 \email{kgerd@ethz.ch}
 \affiliation{%
Rectorate and ETH AI Center, ETH Zurich, 8092 Zurich, Switzerland
}%
\altaffiliation[also at ]{Michigan State University, East Lansing, MI 48823, USA}
\date{\today}

\begin{abstract}
The general availability of general-purpose Large Language Models continues to impact on higher education, yet they may not always be useful for specialized tasks. When using these models, oftentimes the need for particular domain knowledge becomes quickly apparent, and the desire for customized bots arises. Customization holds the promise of leading to more accurate and contextually relevant responses, enhancing the educational experience. This report relates insights and experiences from one particular technical university in Switzerland, ETH Zurich, to describe what ``customizing'' Large Language Models means in practical terms for higher education institutions. 
 \end{abstract}
\maketitle

\section{Large Language Models}
Large Language Models (LLMs) are just one class of artificial intelligence systems, yet they have garnered enormous attention across society, including higher education~\cite{dempere2023impact,vargas2023challenges,balabdaoui2024survey,muscanell2025ai,brodheim2025shaping}. LLMs are driven by numerous parameters, typically billions of learned ``weights.''  Learning here is based on corpora of training materials, which are huge sets of human-produced documents gathered across repositories and the internet; what the system ``knows'' depends on the composition of these materials, where today's general-purpose models aim for a balanced set of general knowledge, optimized for breadth over depth --- a mile wide and an inch deep. In our university context, however, we often require in-depth knowledge within specific domains, which is why there are frequent calls for customized models among academics. This experience report from ETH Zurich aims to illustrate what customization at different levels entails.

At the heart of LLMs is an auto-complete-like mechanism that, given preceding text and these weights, predicts what should come next in the sequence. For illustration, consider a deliberately simplified, whole-word toy example with a tiny training corpus of two sentences:
\begin{itemize}
\item The Alps are a large mountain range in Europe.
\item The Matterhorn is located in the Alps.
\end{itemize}
When given the prompt ``One of the mountains in Europe \ldots,'' such a toy model might continue ``\ldots is the Matterhorn.'' The final result is, ``One of the mountains in Europe is the Matterhorn.''
Note how it draws inferences between the two training sentences: the Matterhorn is in the Alps and the Alps are in Europe. The result is useful and true.

Now consider a different two-sentence training corpus:
\begin{itemize}
\item The Alps are a large mountain range in Europe.
\item The Alps are ideal for skiing.
\end{itemize}
With that corpus, the model has seen the range, the Alps, but no specific mountain names. Given the same prompt, the toy model may somewhat clumsily produce ``\ldots is the Alps.'' The final statement, ``One of the mountains in Europe is the Alps,'' sounds plausible but is incorrect: the Alps are a range, not a mountain. Lacking specific examples, the model falls back to the only salient entity it has seen. In general, the less specific and less relevant the training material, the more the model's completions drift toward improbable or unsupported statements (so-called ``hallucinations''). If the model were human, one might say it does not know what it is talking about.

Real LLMs operate on the level of so-called tokens, which are generally not whole words, but just parts of words, syllables, or punctuation. Our whole-word illustration above also omits important realities of modern LLMs (embeddings, long-range context, calibration, and instruction-following). The point here is narrower: the data the model was trained on strongly influences what it finds ``natural'' to say next, even for simple prompts. Different model families, such as GPT~\cite{radford2018gpt}, Gemini~\cite{gemini2023family}, and LLaMA~\cite{touvron2023llama}, differ in architecture, parameter count, and training corpora. As a rough rule of thumb, larger, more extensively trained models tend to perform better; however, advances in reasoning-oriented training, instruction tuning, and data curation have made size less determinative than it once was.

Performing these inference operations (``running the LLM'')  involves large-scale matrix operations that are usually performed on processing units that were originally designed for graphics operations (GPUs). Given the size of today's models, this is typically the domain of large-scale computing centers --- it would generally not be economically feasible for a single university to run any of the more powerful models 24/7 in a scalable way that would for example sustain hundreds of students in a lecture hall hitting the system at the same time.

Chatbots such as OpenAI's ChatGPT, which is an end-user interface for GPT models, are the most prominent application of LLMs.  However, of possibly equal or higher importance to higher education institutions, are application programming interfaces (APIs), so other campus systems can make use of the LLM capabilities in the background. Those systems may well run on-campus and use LLM-inference as a commodity service.

\section{Customized Large Language Models}
\begin{figure*}
\includegraphics[width=\textwidth]{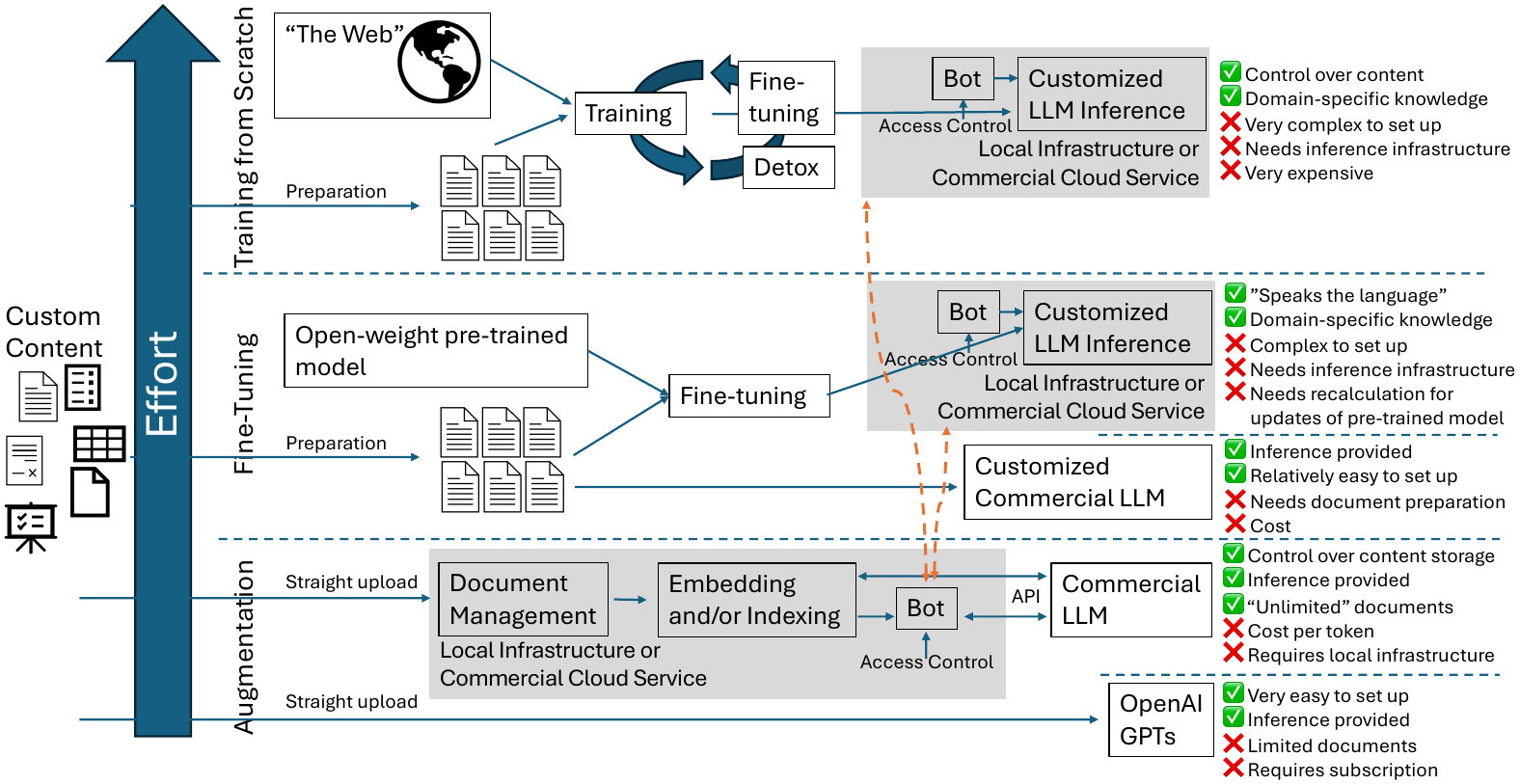}
\caption{Overview of the customization methods. The ``'Effort''-axis can be considered logarithmic.}
\label{fig:options}
\end{figure*}

To customize Large Language Models, there are basically three routes with very different complexities, illustrated in Fig.~\ref{fig:options} and discussed below.

\subsection{Training from Scratch}

By far the most complex and resource-intensive way to get a customized LLM is to build a model from scratch.  Besides having to decide on crucial architectural features, today's models have billions of weight parameters, which require an enormous amount of training data to properly adjust and optimize; the training corpora of today's general-purpose models encompass trillions of tokens. Curating these training corpora is  work-intensive, followed by computationally intensive training over months, followed by supervised and unsupervised tuning and and safety alignment (often called ``detoxification''), which aims to reduce the likelihood of unsafe, unlawful, or harmful outputs and to mitigate the influence of problematic web content. The computing requires considerable resources and power (associated with large CO$_2$ production~\cite{wu2022sustainable}), The process is illustrated in the top row of Fig.~\ref{fig:options}.

\subsubsection{Public Systems}

ETH Zurich (ETHZ) and its sister institution, EPF Lausanne (EPFL), recently released Apertus~\cite{apertus2025}, which is built from scratch by our SwissAI initiative~\cite{swissai_overview_2025}. Designed to be open and fair, it  limited its training corpus to documents permitted for use. As already within its own borders, Switzerland has four official languages, the training corpus also had a higher-than-usual proportion of non-English documents added to the mix. This project was made possible by having access to the high-performance Alps computer~\cite{alps_top500_2025} (not to be confused with the European mountain range that contains the Matterhorn!). Alps is the Swiss National Supercomputing Centre's HPE Cray EX254n system built with NVIDIA GH200 Grace Hopper Superchips, delivering 434.9 PFLOPS for AI- and data-intensive science and ranked \#8 on the June 2025 TOP500 list of most powerful computers --- all this techno-babble to illustrate that training a model from scratch requires computing resources that are usually only available on a national level.

Such a high-performance computer is designed for research, not production purposes. It processes in batches instead of being available 24/7, and one cannot dedicate all its compute power to just one specific application. The day-to-day grind of running the LLM, that is, performing inference, has to be done elsewhere, and that is the next challenge: you need a system that continuously operates at high availability. Without inference infrastructure, training such a model is like building a car in a world that has no roads. You can run it for benchmarks on your own test track, but it never goes anywhere.

Loading and unloading a specific model takes longer than the average user has patience to wait for a reply, so you need to run it ``hot,'' which requires permanently dedicating sufficient GPU resources to absorb peak loads, This is usually the domain of commercial hyperscalers like Microsoft's and Google's data centers. To make their efforts worthwhile and commercially viable, they tend to dedicate their computer resources to the usual suspects of general-purpose, commercial models, and it is hard for a from-scratch, open model to break into this (and if the hyperscalers load such models, it is usually by-the-hour, not per usage: you pay for the time that they dedicate their hardware for your purposes, not for how much it is actually used).

However, besides building another \textit{Large} Language Model, in specialized applications, Small Language Models might do the trick (or even perform better). As the name suggests, these models are much smaller, so they require less compute-power in training, and their inference requirements are far more modest --- in exchange for their knowledge being an inch wide and a mile deep. Coupled with reasoning abilities, these are serious contenders to LLMs in their respective domains.

\subsubsection{Commercial Systems}
Common lore is that powerful commercial systems like GPT-5~\cite{openai_gpt5_systemcard_2025,openai_introducing_gpt5_2025} have essentially ingested the ``whole internet;'' while that may be a hyperbole,  the origin of the training materials is sometimes questionable and the legal situation murky~\cite{frankel2024}. Also,  the manual tuning and detoxification at times employs questionable work practices~\cite{perrigo2023}. As a result, though, their performance is hard to achieve with more ethically built models, and in spite of concerns, many of us rely on their convenience and ever-increasing reliability.

In addition, models like GPT-4o~\cite{gpt4o} , GPT-5, and Gemini are multimodal, so they accept auditory and visual input, and they can produce corresponding output. In the area of higher education, multimodality is particularly interesting for lecture transcriptions (voice and blackboard), as well as handwriting recognition for grading tasks~\cite{kortemeyer2023toward,kortemeyer2024grading}. This generalized multimodality currently is the domain of the largest commercial models; university-generated systems usually implement multimodality only in research settings.

Inference for commercial models is readily available via managed APIs and is typically billed on a usage basis (with some enterprise/education plans also charging per seat; using the earlier analogy, there is plenty of road for these cars). For European universities, major vendors offer contractual terms that exclude customer prompts and completions from model training and provide EU data-residency options or EU processing paths. From an environmental standpoint, shared cloud capacity tends to keep GPUs highly utilized (which reduces power consumption per workload), and large providers increasingly site or procure power near low-carbon grids. For example, ETH Zurich currently uses Swedish data centers via Microsoft Azure contracts.

The trade-off is reduced transparency and control over training-data provenance, data governance, and labor practices in alignment work; institutions effectively purchase availability, cost efficiency, and convenience at the expense of some aspects of digital sovereignty. We will see later how these models can be customized.

\subsection{Fine-Tuning}
Fine-tuning takes an already pre-trained model and modifies the weights. This only works if the original model is ``open-weight,'' meaning, those billions of numbers can be downloaded and locally modified (our Apertus is such an open-weight model; 
while GPT allows for limited fine-tuning (illustrated in the middle row of Fig.~\ref{fig:options}), this has to be done on Open\-AI's platform at considerable cost per token~\cite{gptfine}, and the resulting model needs to be run there). This is similar to editing a pre-written paper: the core content is already in place, and now it is being refined and enhanced. A big advantage is that the model already ``knows'' how to talk and reason, and one only needs to provide the custom data, alas usually after extensive preparation in specialized formats. This requires much less, but still considerable computational effort. After downloading the weights for the pre-trained model, fine-tuning usually requires several iterations; there are common libraries for such efforts, and typical projects make use of the Hugging Face ecosystem~\cite{hugging}.  The process is illustrated in the second-to-the-top row of Fig.~\ref{fig:options}.

While the effort is reasonable, there are caveats. The open-weight models are already carefully pre-trained and tuned, and fine-tuning them can lead to overall worse results: the system may gain factual knowledge, but lose language and reasoning capabilities, and it will likely ``forget'' other things~\cite{he2021analyzing}. In our initial, simplified example, you could add the information that the Matterhorn is in the Alps, but if you are not careful, it may loose the association between the Alps and Europe. Finding the right level of fine-tuning is a balancing act~\cite{lin2023speciality}. It also should be kept in mind that fine-tuned models still ``hallucinate.'' Also, to take advantage of newer versions of the open-weight model, for example going from LLaMa~3.1 to LLaMa~3.2, the fine-tuning would have to be repeated, as all of the weight adjustments would need to be recalculated.

Fine-tuning, however, can help a model better ``speak the language'' of particular disciplines. The most viable route here seems to perform a limited amount of fine-tuning with carefully curated and prepared custom data on university research-computing infrastructure. For example, a research group at ETH Zurich fine-tuned a LLaMa to better speak the language of mathematics tutors~\cite{pal2024autotutor}.

The finished product, unfortunately, has the same inference challenge: where do you run it? Just like with models that are trained from scratch, production requires dedicated hardware, and the more boutique models are being created, the bigger the challenge. There are mechanisms to only load and unload the changes between different fine-tunings of the same pre-trained model, thus reducing wait times when switching models, but that still requires inference infrastructure. Unless the goal is purely research (which of course is legitimate!), any proposal to fine-tune LLMs should include a section where to run the finished product.

\subsection{Augmentation}
Retrieval Augmented Generation (RAG)~\cite{lewis2020retrieval} is a method of customizing a chatbot without changing the LLM. The basic concept is to send relevant reference material alongside the user input: the user submits a prompt to the bot, the bot finds relevant text segments in its database of custom documents, and it then prompts the LLM along the lines of  ``Reply to [{\it user prompt}] using the following background materials: [{\it relevant text segments}].`` For example, if the sentence ``the Matterhorn is located in the Alps'' were part of the text segments sent alongside the user prompt, even the model that out-of-the-box only ``knows'' about the Alps as a skiing place would be able to provide the correct inference. Additional items injected into the prompt or the role may contain instructions of how to deal with situations where no relevant information could be found, and whether or not to have the references that are sent along supersede general knowledge of the model. The setup is illustrated in the second-to-bottom row of Fig.~\ref{fig:options}

For any sufficiently useful custom chatbot, the amount of data in the provided documents is larger than what can be submitted to LLMs due to token limits. Thus, the art is to locate relevant text segments and only send those. A common approach is to convert the provided documents to plain text and then separate those text files into chunks which may correspond to paragraphs or semantically related segments of texts, depending on the algorithm used; this would be like cutting apart a paper into smaller pieces (though, most algorithms include some overlap between the pieces, which would not be possible with paper and scissors).

\begin{figure*}
\includegraphics[width=\textwidth]{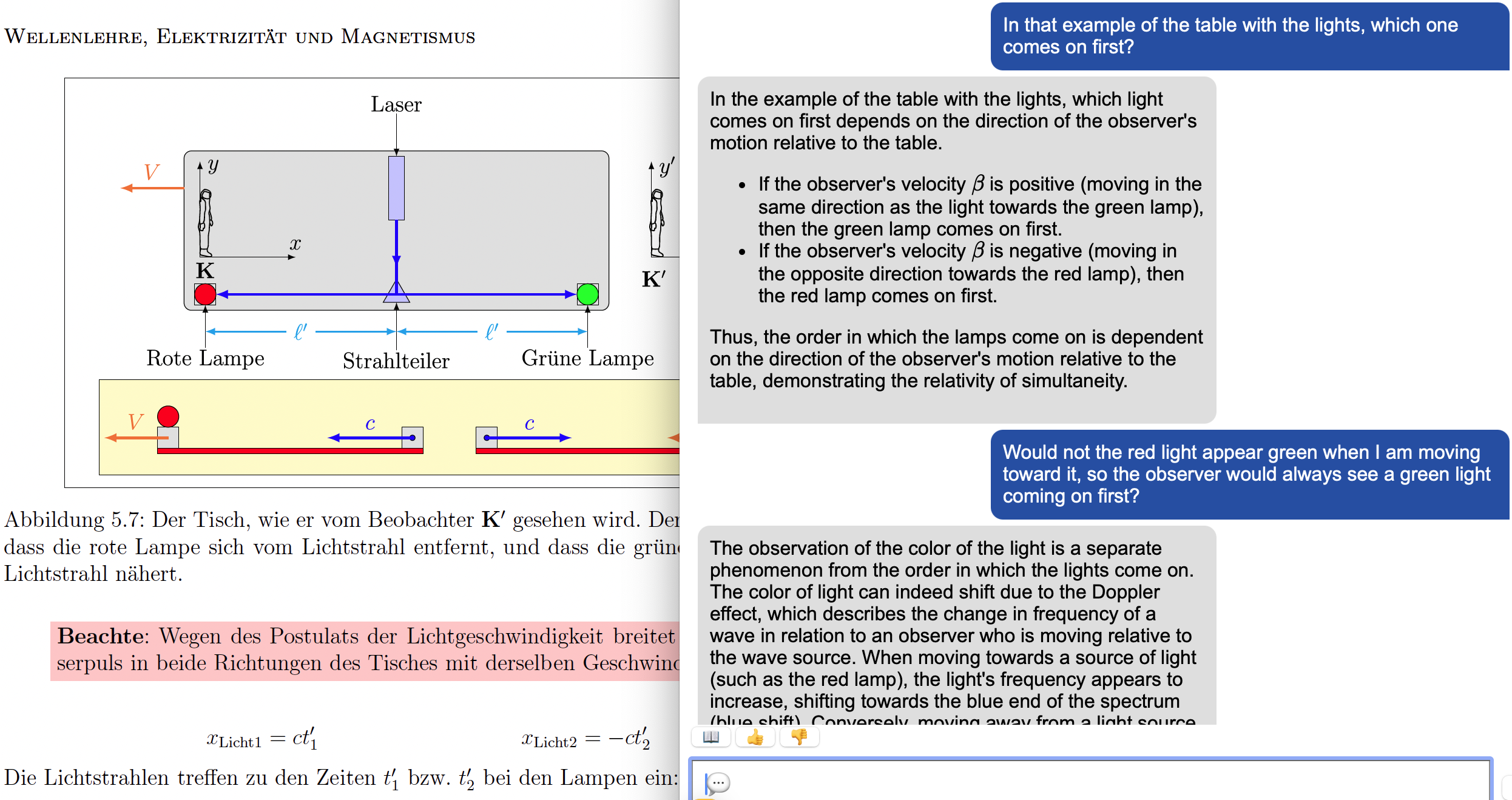}
\caption{Example of the chatbot component of Ethel. Shown on the left is a short excerpt of the lecture script, on the right a dialogue with the system.}
\label{fig:chatbot}
\end{figure*}

The standard method for determining relevance is converting these text chunks into token vectors, using so-called embeddings, for example OpenAI's ada-embeddings~\cite{openaiada}. Embedding is also charged by token, but it turned out that the cost is negligible, even for large document sets. Once the bot is running, the user prompts are also embedded, and the most relevant reference material is identified using cosine-similarity between the text-chunk embedding and the user-prompt-embedding, usually between four and ten chunks. This method often succeeds in finding semantically related text chunks, but it is not a search engine. An alternative approach uses standard indexing search engines to identify chunks, but these struggle with synonyms and when the user prompt is in a different language than the documents (like in Fig.~\ref{fig:chatbot}). A combination of semantic and indexed search is possible.

This approach entails creating a local infrastructure for RAG either through a few hundred lines of code~\cite{ethelrag} on a local low-power server using standard tools like LangChain~\cite{langchain} (the local machine does not do any heavy computing) or buying this service from a cloud provider (``RAG as a service''). For local installations, the service also requires access to a standard LLM via its API as the conversation and reasoning agent~\cite{kortemeyer2024ethel}. Figure~\ref{fig:chatbot} gives an example of a dialogue with such a system.

Local systems can be set up in a day or two, using a dedicated server or VM. Integration into the local IT landscape, for example, setting up access control, can take a little longer.
Code examples are openly available~\cite{ethelrag}.

At higher education institutions, it makes sense to set up one bot instance per course, each one using its own database file or database collection within the shared RAG infrastructure. The one-bot-per-course approach ensures that answers are specific to that course and follow the established notations. Most of all, though, learner questions about Calculus~1 should not be answered using concepts from Calculus~3. Instructors can upload and remove documents for their courses, such as scripts, slides, old exams, exercise sheets; at ETH Zurich, these usually amount to several hundred pages. For the semantic search, these documents need to be embedded, so it usually takes a few minutes until new materials are available in the bot. 

RAG has many advantages, as it is fast to implement and using standard LLMs, so inference is readily available. The system can be ``rewired'' (even on-the-fly) to take advantage of different LLMs or switch to a newer version of the LLM; it can also be used in combination with customized LLMs (illustrated by the orange dashed lines in Fig.~\ref{fig:options}). If prompted system-side to say ``I don't know'' in case the documents provide no relevant information, it is low on hallucinations (but also potentially less useful). It can also be prompted system-side to not directly answer leaner questions, but for example ask follow-up questions and guide the learner toward a solution.

Finally, this architecture can be expanded, and the same materials used to augment other course-specific functionality, such as practice problem generation~\cite{geisler2025ai}, homework feedback~\cite{kortemeyer2024ethel}, or exam grading~\cite{kortemeyer2025assessing,cvengros2025assisting,kortemeyer2025artificial}.

The bottom row of Fig.~\ref{fig:options} shows the simplest customization method, using OpenAI GPTs~\cite{gpts}. The technology is proprietary, but likely RAG. As opposed to local installations, the amount of reference material you can upload is limited, and you have no control on which data center inference is run. A subscription is required to generate GPTs, and unsubscribed users only have limited access to the GPTs before it reverts to a very simple (by today's standards) LLM in the background.

\section{Disclaimer}
This report was put together in the hopes that it may be useful. It can be no more than a snapshot in time, written from the perspective of a technical university on \today{}. The LLM landscape, and in particular the landscape of associated services, is rapidly evolving: systems come and go at a pace that can make this report obsolete within months.

\section{Conclusion}
There is no one-size-fits-all for customization of chatbots. Figure~\ref{fig:options} summarizes the different options. For most purposes, RAG may be sufficient. Careful fine-tuning can help the system ``learn the language'' of its target users and infuse some basic knowledge about the discipline into general pre-trained models, but it requires an order of magnitude more effort to set up and requires a custom inference service. Training a model from scratch is prohibitively complex and expensive in all but a few special situations.

\begin{acknowledgments}
The author would like to thank Imanol Schlag for proofreading and critiquing the manuscript.
\end{acknowledgments}

\bibliography{custombots}
\end{document}